\documentclass[twoside,onecolumn,english]{revtex4-2}

\usepackage[T1]{fontenc}          
\usepackage[latin9]{inputenc}     

\usepackage{babel}
\AtBeginDocument{}

\usepackage{amsmath}
\usepackage{amssymb}

\usepackage{bm}

\usepackage{graphicx}
\makeatletter
\newcommand{\lyxmathsym}[1]{%
  \ifmmode
    \begingroup
      \def\b@ld{bold}%
      \text{\ifx\math@version\b@ld\bfseries\fi#1}%
    \endgroup
  \else
    #1%
  \fi
}
\makeatother

\begin{document}
\title{Complexity and Information in Quantum and Classical Trajectories}
\author{Hira Ali}
\affiliation{Department of Physics, University of Swabi, Khyber Pakhtunkhwa, Pakistan}
\author{Naeem Shahid}
\affiliation{Department of Physics, Pennsylvania State University, New Kensington,
PA 15068, USA.}
\begin{abstract}
We analyze emission trajectories from a driven--dissipative two--qubit system and a classical telegraph model with matched rates. Using Lempel--Ziv complexity, mutual information, and temporal correlations, we show that both models undergo a transition from independent to synchronized dynamics as coupling increases, but only the quantum trajectories develop enhanced complexity and sustained information sharing at large drive--to--decay ratio. Classical correlations are short--lived and quickly suppressed by strong drive. A strong complexity--information correlation appears uniquely in the quantum case, providing a clear trajectory--level signature of quantum effects. These results show that complexity and information measures extracted directly from jump records provide an efficient way to distinguish quantum and classical dynamics in open systems.
\end{abstract}
\maketitle

\section{Introduction}

A driven system coupled to an external bath can generate rich dynamical structures at the boundary between complete randomness and perfect order, often referred to as complex dynamics. The emergence of such complexity is a central focus in open quantum systems \cite{Carmichael1993,Aolita:2015aa,Bhattacharyya:2022aa,Liu:2023aa,Estrada2024,Rivat2024,Monteiro_2025}. In quantum spin systems, these features arise not only from the interplay between coherent drive and stochastic jumps, but also through quantum effects such as entanglement and measurement backaction \cite{Wiseman2010,Xu:2010aa,Franco:2012aa,Xu_2013,Gessner:2014aa}.

One way to probe these dynamics is through the unraveling of the Lindblad master equation into quantum jump trajectories \cite{Dalibard1992,Gisin_1992,Carmichael1993,M_lmer_1993,Wiseman_1996,Plenio:1998aa,Zoller2004,Daley_2014,Parthasarathy:2017aa}. Each trajectory serves as a footprint of a possible measurement history, shaped jointly by coherent evolution and noisy decay processes.

Recent studies have explored driven--dissipative spin models and uncovered collective phenomena and nonequilibrium phases, including bistability in spin--cavity systems using a Maxwell--Bloch approach \cite{Krimer:2019aa}, dissipation-induced bistability in the Dicke model \cite{Garbe_2020}, metastable behavior in interacting spin ensembles \cite{Rose_2016}, and nonequilibrium critical fixed points in dissipative Ising chains \cite{Young:2020aa}. Similar phases with critical slowing down have also been identified in driven Bose--Hubbard lattices, which are not strictly spin models but share the same nonequilibrium transition physics \cite{Vicentini:2018aa}.

On the classical side, stochastic models such as coupled telegraph processes and noise--driven oscillators have provided additional insight into synchronization, noise--induced ordering, and collective switching phenomena \cite{Kim:1987aa,McNamara:1989aa,Kiss:2005aa,Saira:2007aa,Wold:2012aa}. Comparing quantum models with their classical analogs therefore offers a natural framework for identifying quantum signatures in correlated emission dynamics.

The use of quantum emission trajectories as a signature for underlying correlations and irreversibility has also become increasingly central in recent years \cite{Garrahan:2010aa,Chantasri:2016aa,Carollo:2019aa,Carollo2021,Vu:2021aa,Ferrari:2025aa}. Trajectory--level statistics, linked to entropy production, dynamical phase transitions, and computational complexity, provide a complementary perspective on open quantum dynamics that goes beyond steady--state observables.

In this work, we construct and compare a two--qubit Ising--like model and its classical stochastic analog based on interacting telegraph processes. By analyzing emission trajectories, mutual information, and temporal correlations, we quantify how the coupling strength $J$ controls the transition from independent to synchronized dynamics. Our results show that the quantum model develops enhanced correlations and information content at intermediate coupling, where coherence competes with dissipation and stochastic jumps to generate the largest separation from the classical model.

Our approach builds on established frameworks of quantum--jump unravelings \cite{Dalibard1992,Carmichael1993}, driven--dissipative spin models \cite{Buca2019,Xu:2019aa}, and interacting stochastic processes \cite{McNamara:1989aa,Kiss:2005aa}. The analysis of correlations and mutual information in emission records connects to recent work on quantum complexity and dynamical irreversibility \cite{Lesanovsky:2013aa,Manzano:2019aa,Gneiting:2021aa,Radaelli:2024aa}.

\section{Model and Effective Dynamics}

We consider two coupled two--level systems (qubits), each driven coherently and coupled to independent Markovian decay channels. The system Hamiltonian in the rotating frame is,

\begin{equation}
H=\frac{\Omega}{2}\left(\sigma_{x}^{(1)}+\sigma_{x}^{(2)}\right)+J\,\sigma_{z}^{(1)}\sigma_{z}^{(2)},\label{quantum_H}
\end{equation}

where $\Omega$ is the Rabi drive amplitude and $J$ denotes the Ising--type coupling strength. 

Each qubit undergoes spontaneous emission via local jump operators,

\begin{equation}
L_{1}=\sqrt{\gamma}\,\sigma_{-}^{(1)},\qquad L_{2}=\sqrt{\gamma}\,\sigma_{-}^{(2)},\label{eq:jump_ops},
\end{equation}

with decay rate $\gamma$.

Between quantum jumps, the state evolves under the non--Hermitian effective Hamiltonian,

\begin{equation}
H_{\mathrm{eff}}  =  H-\frac{i}{2}\sum_{i=1,2}L_{i}^{\dagger}L_{i}\label{eq:H_eff}.
\end{equation}

The corresponding short--time propagator is,

\begin{equation}
U_{\mathrm{eff}}=e^{-iH_{\mathrm{eff}}\,dt}.\label{eq:U_evolve}
\end{equation}

At each time step $dt$, the normalized wavefunction $|\psi(t)\rangle$ evolves as,

\begin{equation}
	|\psi(t+dt)\rangle = 
	\begin{cases}
		{\displaystyle 
			\frac{L_{i}|\psi(t)\rangle}{\|L_{i}|\psi(t)\rangle\|},} 
		& \text{\begin{tabular}[t]{@{}l@{}}
				with probability\\
				$p_{i} = \gamma\,dt\,\langle\psi(t)|\sigma_{+}^{(i)}\sigma_{-}^{(i)}|\psi(t)\rangle,$
		\end{tabular}}\\[1.2em]
		{\displaystyle 
			\frac{U_{\mathrm{eff}}|\psi(t)\rangle}{\|U_{\mathrm{eff}}|\psi(t)\rangle\|},} 
		& \text{otherwise.}
	\end{cases}
	\label{eq:psi_dt}.
\end{equation}

We also introduce a minimal classical analog, an interacting telegraph--spin model (ITSM). Each spin $s_{i}(t)\in\{0,1\}$ flips stochastically with a base rate determined by effective parameters $\Omega_{\mathrm{eff}},\beta J$ and $\gamma$. The instantaneous flip probability is,

\begin{equation}
p_{i}=dt\gamma \Omega_{\mathrm{eff}}e^{-\beta J(2s_{i}-1)(2s_{j}-1)},\label{eq:pflip}
\end{equation}

where $s_{j}$ denotes the partner spin.

This exponential bias favors aligned or anti-aligned configurations depending on the sign of $\beta J$ , mimicking effective thermal interactions. The parameter $\beta$ is treated as a scaling factor for our purpose and we will keep it fixed, in particular $\beta=1$.

We will record the jump events for both models as a binary sequence of emissions with elements $r_{i}(t)\in\{0,1\}$ from each emitter $i=1,2$. One such sequence over time is a stochastic trajectory. The ensemble of these trajectories encodes both dynamical correlations and quantum information sharing between the subsystems. 

From these trajectories, we compute autocorrelations $C_{ii}(\tau)=\langle r_{i}(t)r_{i}(t+\tau)\rangle_{t}$, cross-correlations $C_{12}(\tau)=\langle r_{1}(t)r_{2}(t+\tau)\rangle_{t}$ to measure emission synchronization. We will also use Lempel--Ziv (LZ) complexity to quantify the information content or unpredictability of a sequence \cite{Ziv_1977, Lempel_1976}. 

We will also be interested in mutual information $I_{AB}(t)=S(\rho_{A})+S(\rho_{B})-S(\rho_{AB})$ from the instantaneous state $\rho(t)=\langle\psi(t)|\psi(t)\rangle$, with $S(\rho)=-\mathrm{Tr}[\rho\ln\rho]$ the von Neumann entropy. Finally, classical mutual information $I_{\mathrm{cl}}=H(s_{1})+H(s_{2})-H(s_{1},s_{2}),$ with $H(x)=-\sum_{x}P(x)\ln P(x)$ as the Shannon entropy, will also be computed.

\section{Trajectory--Level Behavior}

\begin{figure*}[t]
	\includegraphics[width=1\textwidth]{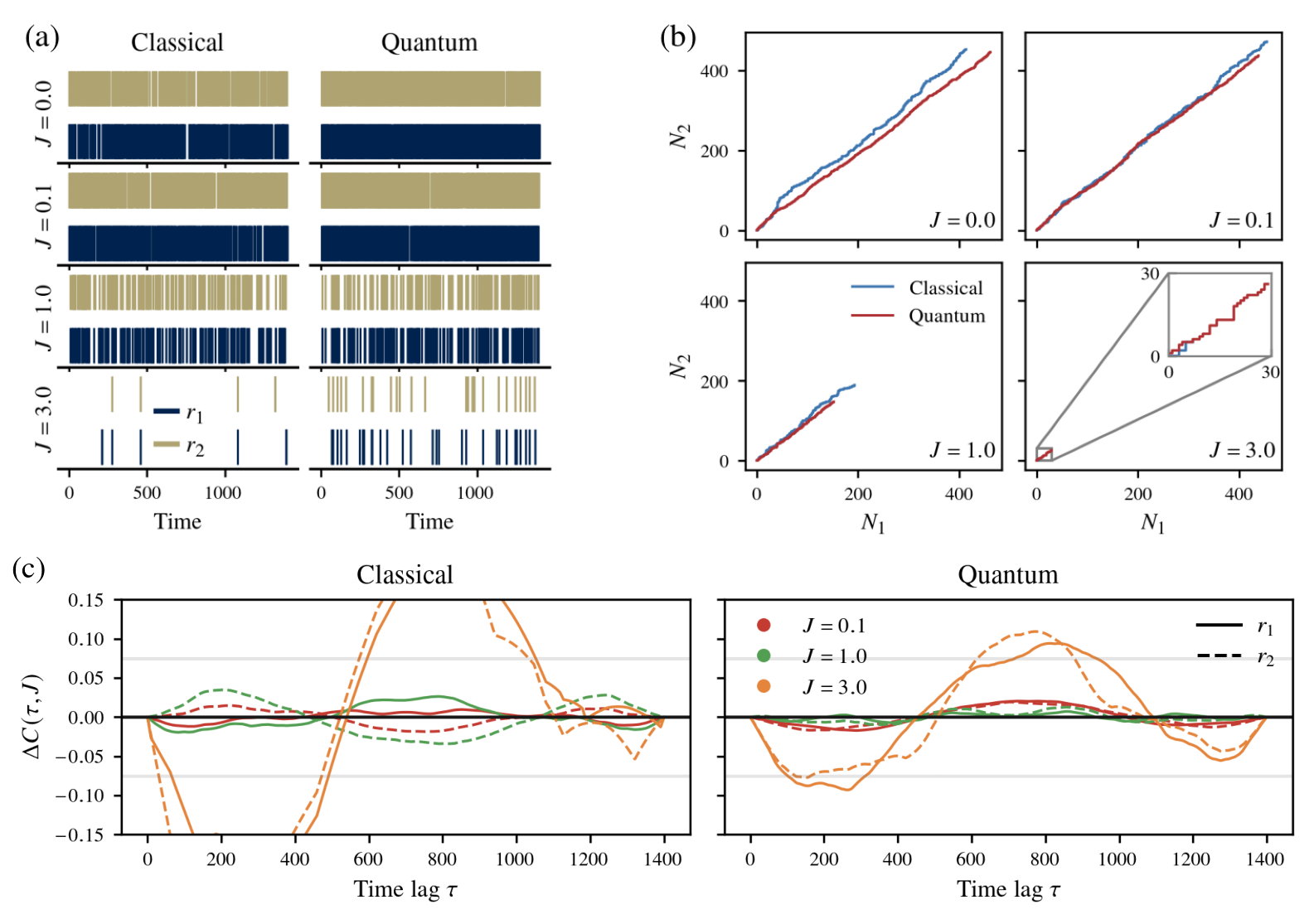}
	\caption{Trajectory--level comparison of the classical interacting telegraph model and the quantum jump model for $\Omega=1.0$, and $\gamma=1.0$. (a) Representative emission trajectories for increasing coupling strength $J$. For weak coupling, both systems produce independent jump records. At intermediate $J$, intermittent bursts and correlated intervals appear, marking the onset of synchronized dynamics. For large $J$, the classical model locks into a frozen configuration, while the quantum model retains residual fluctuations due to coherence and measurement backaction. (b) Cumulative emission counts $N_{2}$ versus $N_{1}$. Weak coupling produces near--diagonal behavior consistent with independent activity. Increasing $J$ bends the curves into step--like structures, reflecting co--jumps and partial locking. Quantum trajectories remain smoother and less abrupt than the classical ones, showing the effect of quantum coherence in suppressing sharp switching. (c) Autocorrelation differences $\Delta C(\tau,J)=C(\tau,J)-C(\tau,0)$. Classical trajectories (left) develop slow, large--amplitude oscillations as $J$ increases, while quantum trajectories (right) remain more bounded and regular.}
	\label{fig:single_traj}
\end{figure*}

Fig.~\ref{fig:single_traj}(a) shows the effect of coupling strength $J$ on single trajectories for both the quantum and classical models. Both emission channels remain uncoupled when the interaction is weak, which is to be expected. As $J$ approaches intermediate values, we observe intermittent bursts of emissions separated by quiet intervals. This regime marks the onset of synchronized dynamics, which later leads to a decrease in complexity and an increase in mutual information (see Sec.\ref{sec:complexity_Info}). Once the coupling strength exceeds its intermediate range, the trajectories effectively freeze, and both systems settle into a steady, correlated configuration.

Overall, the qualitative similarities between the two columns indicate that the classical model captures the same dynamical transitions as the quantum one across all values of $J$.

Fig.~\ref{fig:single_traj}(b) presents cumulative jump counts for each subsystem in both models. In the weak--coupling regime, the curves have nearly identical slopes, indicating independent emission records fluctuating around the diagonal.

As $J$ increases, the curvature and local slope begin to change, signaling the onset of partial synchronization. This eventually develops into stepwise co--jumping for strong coupling ($J=3$). The total number of jumps grows slower in this regime, causing the curves to terminate earlier than in the weakly coupled case. A small offset between the quantum and classical traces persists. Quantum trajectories show delayed and smoothed joint events due to residual coherence, while the classical model exhibits sharper, more discrete switching.

The role of coupling can be further resolved by examining the temporal autocorrelation function $C_{ii}(\tau)=\langle n_i(t),n_i(t+\tau)\rangle_t$ for different values of $J$. Fig.~\ref{fig:single_traj}(c) shows the difference $\Delta C(\tau,J)=C(\tau,J)-C(\tau,0)$ relative to the uncoupled baseline. Near $J=0$, both systems show almost no temporal correlation, consistent with independent emissions. As $J$ increases, the classical system exhibits slowly varying oscillations with larger amplitude, while the quantum system shows smoother, more bounded oscillations.

These trends highlight a key distinction, the quantum system retains short--term memory through residual coherence, while the classical system displays long--range coherence driven purely by stochastic synchronization.

\begin{figure}
	\includegraphics[width=1\columnwidth]{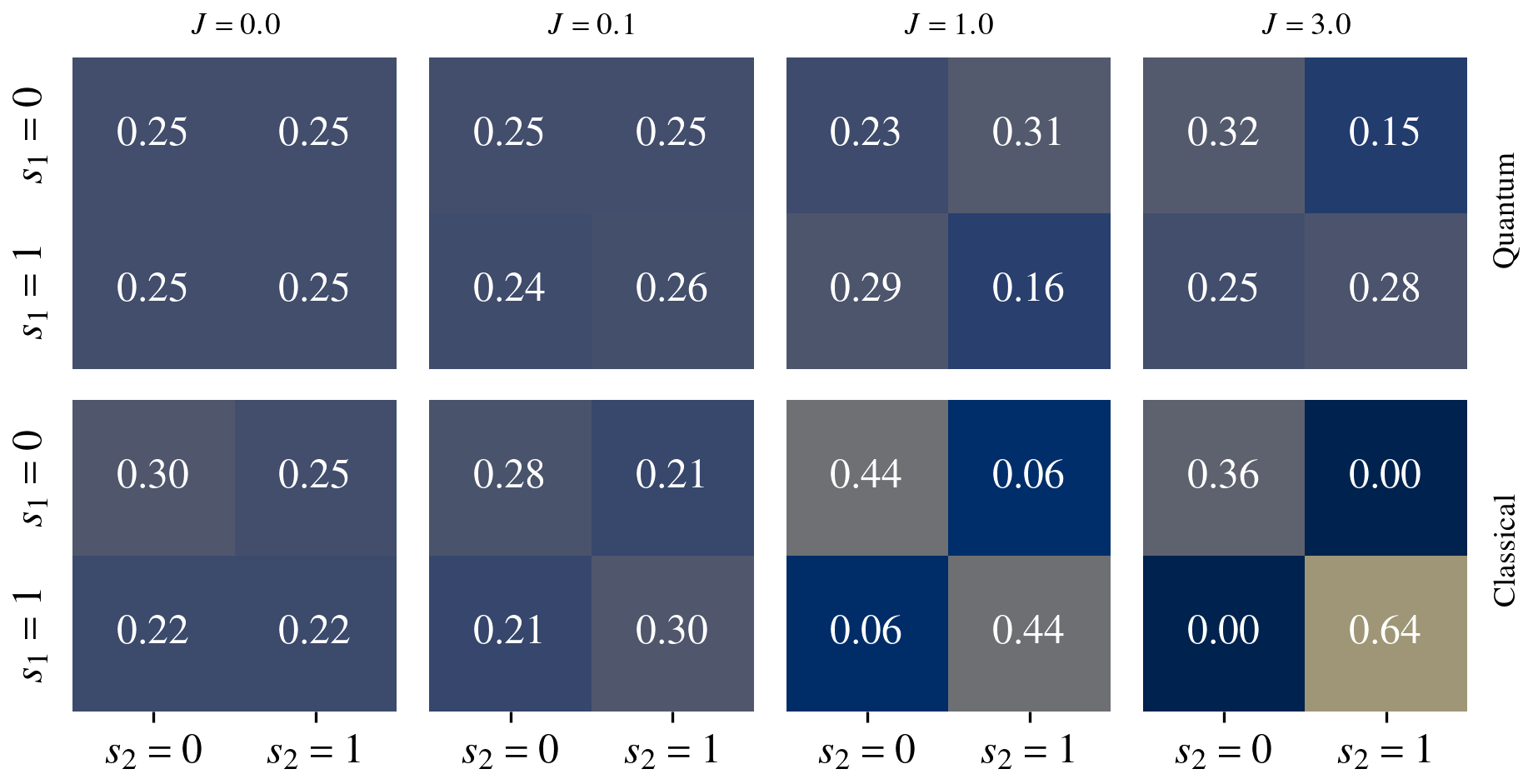}
	\caption{Joint--state occupancy distributions for quantum and classical dynamics with the same parameters as in Fig.~\ref{fig:single_traj}. Heatmaps show the joint probabilities $P(s_{1},s_{2})$ for $J=0,0.1,1$, and 3. The top row displays quantum and the bottom row shows classical results. At weak coupling, both models exhibit nearly uniform occupancy. Increasing $J$ enhances correlations and population asymmetry. For large $J$, classical trajectories collapse into a single dominant state, whereas the quantum system continues to explore all states through coherent fluctuations.}
	\label{fig:occupancy}
	
\end{figure}

Finally, we compute joint--state occupancy probabilities from long--time trajectories. As shown in Fig.~\ref{fig:occupancy}, both systems exhibit nearly uniform occupation probabilities at weak coupling, consistent with uncorrelated single--spin dynamics. At intermediate coupling ($J=1$), the classical system begins to populate the diagonal entries (states 00 and 11), while the quantum system remains nearly uniform. 

In the strong--coupling regime ($J=3$), the classical system collapses into a nearly frozen configuration along the diagonal, reflecting a locked state. However, the quantum system continues to explore all four states due to residual coherence. This highlights another key difference: the classical model locks into synchronized configurations, whereas the quantum system preserves probabilistic exploration even under strong coupling.

\section{\label{sec:complexity_Info}Complexity and Information Measures}

\subsection{Lempel--Ziv complexity trends at $J=0$}

In the uncoupled limit ($J=0$), the dynamics of the two qubits/bits reduce to those of a single driven two--level system. For a fair comparison, both the quantum model and its classical toy version must use the same effective steady--state rates. Enforcing this condition gives the effective classical drive,
\begin{equation}
	\Omega_{\rm eff}=\frac{\gamma\Omega^{2}}{\gamma^{2}+2\big(\Omega^{2}+J^{2}\big)}.
	\label{mapping}
\end{equation}
The $J$ term in the above equation captures the crossover from overdamped behavior to underdamped oscillations, which we will discuss later in this section.

\begin{figure*}[t]
	\includegraphics[width=1\textwidth]{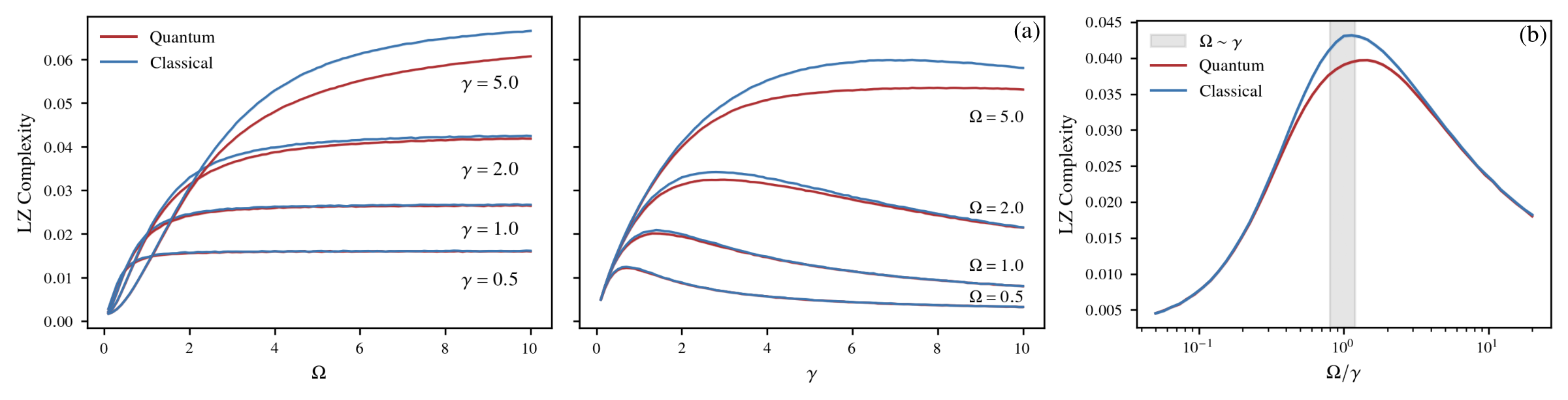}
	\caption{Lempel--Ziv complexity of joint jump sequences in the uncoupled limit ($J=0$). (a) Complexity versus drive strength $\Omega$ at fixed decay rates $\gamma$. Increasing $\Omega$ enhances temporal structure, producing a complexity peak near $\Omega\sim\gamma$ before saturating at large drive. (b) Complexity versus decay rate $\gamma$ at fixed $\Omega$, showing the same crossover from coherent to incoherent emission approached from the opposite direction. (c) LZ complexity plotted against the ratio $\Omega/\gamma$ using logarithmic sampling. Both models collapse onto the same curve with a clear peak at $\Omega/\gamma\approx 1-2$, identifying the boundary between rate--limited and coherence--limited dynamics. Quantum curves lie slightly below classical ones at intermediate parameters due to measurement backaction. Peak LZ values: quantum 0.0257 $\pm$ 0.0018, classical 0.0248 $\pm$ 0.0016; $t =$ 0.375, $p =$ 0.71.}
	\label{fig:map_omega}
\end{figure*}

Figure~\ref{fig:map_omega}(a, left panel) shows how the joint LZ complexity changes with drive strength $\Omega$ at fixed $\gamma$. For small $\gamma$ (e.g., $\gamma=0.5$), both quantum and classical trajectories produce more structured and regular patterns as the drive increases. Around $\Omega\sim\gamma$, the system sits at the boundary between coherent and incoherent dynamics, producing partially ordered sequences and local complexity peaks. For $\Omega \gg \gamma$, oscillations stabilize and the complexity saturates.

Figure~\ref{fig:map_omega}(a, right panel) shows the complementary plot in which $\gamma$ is varied at fixed $\Omega$. Increasing $\gamma$ initially increases irregularity in the jump record, but strong decay eventually suppresses structure and drives both models toward Poisson--like behavior, lowering the complexity. This is the same crossover seen in panel (a), just approached from the opposite direction.

Quantum and classical curves remain very close across all parameters, but small deviations appear at moderate $\Omega$ or $\gamma$. Quantum trajectories have slightly lower complexity due to measurement backaction and short--range temporal correlations, which reduce apparent randomness that do not exist in the classical model.

Finally, Fig.~\ref{fig:map_omega}(b) plots the complexity directly against the ratio $\Omega/\gamma$ using logarithmic sampling. Both models collapse onto the same characteristic curve with a clear complexity peak near $\Omega/\gamma \approx 1-2$, marking the transition from rate--limited to coherence--limited dynamics. The peak values for the quantum and classical cases agree within statistical uncertainty, 0.0257 $\pm$ 0.0018 (quantum) versus 0.0248 $\pm$ 0.0016 (classical), yielding $t =$ 0.375 and $p =$ 0.71.

\subsection{Complexity, mutual information and complexity--information correlation}

\begin{figure*}[t]
	\includegraphics[width=1\textwidth]{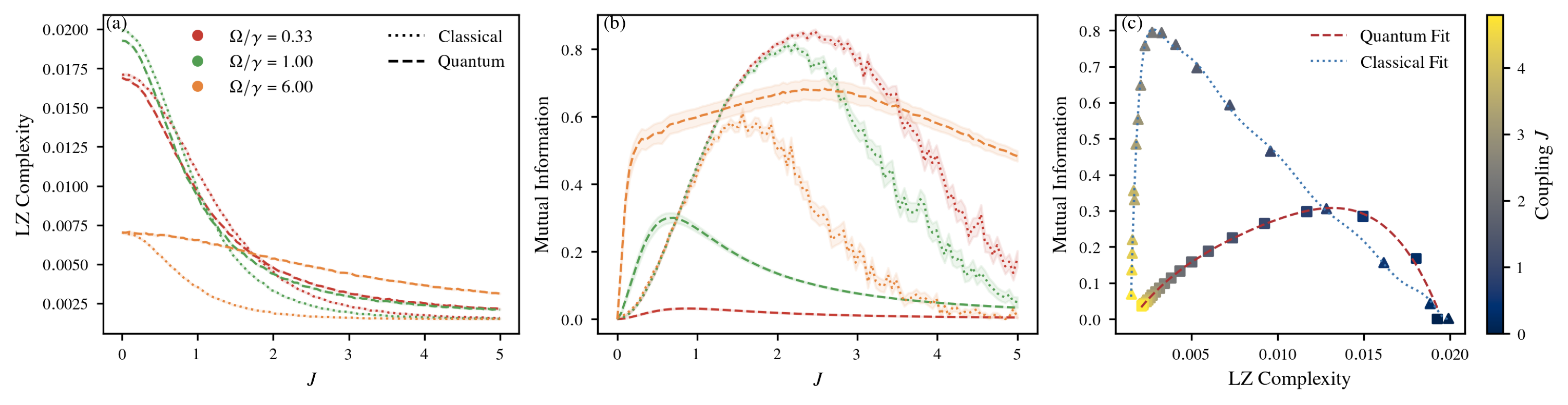}
	\caption{(a) Lempel?Ziv (LZ) complexity versus coupling strength $J$ for quantum (dashed) and classical (dotted) models for different drive--to--decay ratios $\Omega/\gamma$. Increasing coupling suppresses randomness and drives both systems toward structured, synchronized dynamics, with the quantum model retaining higher complexity at large $\Omega/\gamma$. (b) Mutual information versus $J$ for the same parameters. Quantum mutual information increases with $\Omega/\gamma$ and peaks at intermediate coupling, reflecting coherence--assisted information exchange, while classical mutual information peaks at smaller $\Omega/\gamma$ and decreases at strong coupling due to stochastic synchronization and freezing.
	(c) Mutual information plotted against LZ complexity, with points colored by coupling strength $J$. The quantum model exhibits a strong monotonic correlation between information and complexity, whereas the classical model shows a weaker, non--monotonic relation, indicating fundamentally different mechanisms for correlation generation. Spearman coefficient: quantum $\rho = 0.817$, $p = 4.0 \times 10^{-25}$; classical  $\rho = 0.162$, $p = 0.11$.}
	\label{fig:lz_info_all}
\end{figure*}

Fig.~\ref{fig:lz_info_all}(a) shows LZ complexity and mutual information as a function of coupling strength $J$ for several values of the drive--to--decay ratio $\Omega/\gamma$. Overall, both systems display increasingly structured and less random dynamics as $J$ increases, reflected in the monotonic decrease of LZ complexity.

For small $\Omega/\gamma$, dissipation dominates, leading to a rapid fall in complexity compared to the larger ratio ($\Omega/\gamma = 6$), where the dynamics are drive--dominated and can sustain fluctuations even for relatively large $J$. The difference between quantum and classical trends becomes more visible at higher drive strength, where quantum coherence begins to influence the dynamics. In the quantum model, coherence combined with stochastic jumps generates nontrivial temporal correlations that keep the complexity higher than in the classical case. In contrast, the classical model that lacks coherence, the coupling $J$ mainly enforces synchronization through probabilistic flips, eventually producing simple, repetitive structures and thus lower complexity.

The mutual information trends (Fig.~\ref{fig:lz_info_all}(b)) show a much stronger separation between the quantum and classical models. For small $\Omega/\gamma$, the two qubits in the quantum model remain almost independent, and essentially no information is exchanged for any $J$. As $\Omega/\gamma$ increases, coherence builds up, allowing the qubits to communicate. Coupling $J$ then drives the system into a synchronized regime where mutual information increases and eventually saturates (most clearly around $\Omega \sim \gamma$ and $J \sim 1-2$). Beyond this point, further increases in $J$ suppress additional fluctuations, and the mutual information drops as the trajectories freeze. For large $\Omega/\gamma$ (e.g., 6), strong correlations appear even at small $J$ and remain nearly constant over a broad range before eventually decreasing at strong coupling.

The classical model shows broadly similar qualitative features, but the role of $\Omega/\gamma$ is completely different. Since there is no coherence or superposition, the dependence on drive strength is purely stochastic and comes from the modified probability factor that governs neighbor--induced flips. As a result, mutual information peaks at a smaller $\Omega/\gamma$ than in the quantum case. A stronger drive in the classical model simply randomizes the jumps and washes out any correlations that $J$ tries to build. The synchronization created by $J$ is therefore short--lived, and once mutual information reaches its maximum, it decreases steadily as $J$ increases further.

Fig.~\ref{fig:lz_info_all}(c) shows the direct correlation between Lempel--Ziv complexity and mutual information. The Spearman coefficient for the quantum model, $\rho = 0.817$ (with $p = 4.0 \times 10^{-25}$), indicates a highly significant monotonic relationship. This demonstrates that quantum dynamics generate correlated yet nonrandom patterns, where informational richness is tightly linked to dynamical complexity.

In contrast, the classical model exhibits mutual information purely through stochastic synchronization, and the much weaker correlation ($\rho = 0.162$, $p = 0.11$) indicates that complexity and information content remain largely independent. Classical trajectories may become complex without increasing information exchange, reflecting the absence of coherence and feedback--driven structure.

\section{Conclusion and Outlook}

In this work, we analyzed quantum signatures in emission trajectories generated by a coupled qubit system compared with a classical stochastic analogue based on interacting telegraph processes. Both models were driven and dissipative with identical rates. Their emission records show that the classical system reproduces the broad dynamical behavior of the quantum one as the drive--to--decay ratio ($\Omega/\gamma$) and coupling ($J$) are varied.

The LZ complexity, however, clearly separates the two models. The quantum system generates richer and more structured patterns due to coherence and measurement backaction, especially at large $\Omega/\gamma$. These features sustain even at strong coupling, while the classical trajectories collapse into a frozen or locked state as $J$ increases.

The information shared between the two subsystems in each model gives a clear signal of strong quantum effects. At large $\Omega/\gamma$, the quantum model develops significantly higher mutual information, supported by coherence and enhanced by coupling. In particular, the intermediate values of $J$ mark the region of interplay between coherence, dissipation, and stochastic jumps dynamics, where quantum trajectories explore structured patterns that the classical model cannot reproduce. In the classical case, the short--term information is entirely exchanged through coupling $J$ provided that the drive--induced randomization is weak. A strong drive increases classical complexity but simultaneously washes out the short--term correlations created through coupling.

The strong Spearman correlation between LZ complexity and mutual information in the quantum case further indicates that information sharing and dynamical structure grow together, in contrast to the classical model where both quantities evolve largely independently.

Overall, these results show that trajectory--level statistics provide a compact and experimentally accessible framework for identifying quantum signatures in open systems. Of course, the analysis can be extended to larger networks, multi--mode systems, and non--Markovian environments. Here we focused on the simplest setting to highlight how complexity and information flow behave at the trajectory level, leaving these broader questions for future work.

\acknowledgments
We would like to thank Mellissia Richards for reading and commenting on the manuscript and Farhan Saif for pointing out some clarifications in a plot. 

\bibliographystyle{unsrt}
\bibliography{complexity}

@article{Chantasri:2016aa,
	author = {Areeya Chantasri},
	doi = {10.1103/PhysRevX.6.041052},
	journal = {Physical Review X},
	number = {4},
	title = {Quantum Trajectories and Their Statistics for Remotely Entangled Quantum Bits},
	url = {https://doi.org/10.1103/PhysRevX.6.041052},
	volume = {6},
	year = {2016}}

@article{Ferrari:2025aa,
	author = {Filippo Ferrari},
	doi = {10.1103/PhysRevResearch.7.013276},
	journal = {Physical Review Research},
	number = {1},
	title = {Dissipative quantum chaos unveiled by stochastic quantum trajectories},
	url = {https://doi.org/10.1103/PhysRevResearch.7.013276},
	volume = {7},
	year = {2025}}

@article{Carollo2021,
	author = {Carollo, Federico and Garrahan, Juan P. and Jack, Robert L.},
	day = {09},
	doi = {10.1007/s10955-021-02799-x},
	issn = {1572-9613},
	journal = {Journal of Statistical Physics},
	month = {Jul},
	number = {1},
	pages = {13},
	title = {Large Deviations at Level 2.5 for Markovian Open Quantum Systems: Quantum Jumps and Quantum State Diffusion},
	url = {https://link.springer.com/content/pdf/10.1007/s10955-021-02799-x.pdf},
	volume = {184},
	year = {2021}}

@article{Carollo:2019aa,
	author = {Federico Carollo and Robert L. Jack and Juan P. Garrahan},
	doi = {https://doi.org/10.1103/PhysRevLett.122.130605},
	eprint = {1811.04969},
	journal = {Phys. Rev. Lett.},
	pages = {130605},
	title = {Unravelling the large deviation statistics of Markovian open quantum systems},
	url = {https://arxiv.org/pdf/1811.04969.pdf},
	volume = {122},
	year = {2019}}

@article{Saira:2007aa,
	author = {Olli-Pentti Saira and Ville Bergholm and Teemu Ojanen and Mikko Mottonen},
	doi = {https://doi.org/10.1103/PhysRevA.75.012308},
	eprint = {quant-ph/0605241},
	journal = {Phys. Rev. A},
	pages = {012308},
	title = {Equivalent qubit dynamics under classical and quantum noise},
	url = {https://arxiv.org/pdf/quant-ph/0605241.pdf},
	volume = {75},
	year = {2007}}

@article{Wold:2012aa,
	author = {Henry J. Wold and H{\aa}kon Brox and Yuri M. Galperin and Joakim Bergli},
	doi = {https://doi.org/10.1103/PhysRevB.86.205404},
	eprint = {1206.2174},
	month = {06},
	title = {Decoherence of a qubit due to a quantum fluctuator or to a classical telegraph noise},
	url = {https://arxiv.org/pdf/1206.2174.pdf},
	year = {2012}}

@article{Kim:1987aa,
	author = {M. S. Kim},
	doi = {10.1103/PhysRevA.36.5265},
	journal = {Physical Review A},
	number = {11},
	pages = {5265--5270},
	title = {Quantum-jump telegraph noise and macroscopic intensity fluctuations},
	url = {https://doi.org/10.1103/PhysRevA.36.5265},
	volume = {36},
	year = {1987}}

@book{Zoller2004,
	author = {C. W. Gardiner and Peter Zoller},
	edition = {3rd},
	number = {978-3-540-22301-6},
	publisher = {Springer},
	title = {Quantum Noise: A Handbook of Markovian and Non-Markovian Quantum Stochastic Methods with Applications to Quantum Optics},
	year = {2004}}

@article{Parthasarathy:2017aa,
	author = {K. R. Parthasarathy and A. R. Usha Devi},
	doi = {https://doi.org/10.1063/1.4998714},
	eprint = {1705.00520},
	journal = {Journal of Mathematical Physics},
	pages = {082204},
	title = {From quantum stochastic differential equations to Gisin-Percival state diffusion},
	url = {https://arxiv.org/pdf/1705.00520.pdf},
	volume = {58},
	year = {2017}}

@article{Plenio:1998aa,
	author = {M. B. Plenio},
	doi = {10.1103/RevModPhys.70.101},
	journal = {Reviews of Modern Physics},
	number = {1},
	pages = {101--144},
	title = {The quantum-jump approach to dissipative dynamics in quantum optics},
	url = {https://doi.org/10.1103/RevModPhys.70.101},
	volume = {70},
	year = {1998}}

@article{Gisin_1992,
	author = {Gisin, N and Percival, I C},
	doi = {10.1088/0305-4470/25/21/023},
	issn = {1361-6447},
	journal = {Journal of Physics A: Mathematical and General},
	month = nov,
	number = {21},
	pages = {5677--5691},
	publisher = {IOP Publishing},
	title = {The quantum-state diffusion model applied to open systems},
	url = {http://dx.doi.org/10.1088/0305-4470/25/21/023},
	volume = {25},
	year = {1992}}

@article{M_lmer_1993,
	author = {M{\o}lmer, Klaus and Castin, Yvan and Dalibard, Jean},
	doi = {10.1364/josab.10.000524},
	issn = {1520-8540},
	journal = {Journal of the Optical Society of America B},
	month = mar,
	number = {3},
	pages = {524},
	publisher = {Optica Publishing Group},
	title = {Monte Carlo wave-function method in quantum optics},
	url = {http://dx.doi.org/10.1364/JOSAB.10.000524},
	volume = {10},
	year = {1993}}

@article{Gessner:2014aa,
	author = {M. Gessner and M. Ramm and T. Pruttivarasin and A. Buchleitner and H.-P. Breuer and H. Haeffner},
	doi = {https://doi.org/10.1038/nphys2829},
	eprint = {1311.4489},
	journal = {Nature Physics},
	pages = {105},
	title = {Local Detection of Quantum Correlations with a Single Trapped Ion},
	url = {https://arxiv.org/pdf/1311.4489.pdf},
	volume = {10},
	year = {2014}}

@article{Xu:2010aa,
	author = {Jin-Shi Xu and Chuan-Feng Li and Cheng-Jie Zhang and Xiao-Ye Xu and Yong-Sheng Zhang and Guang-Can Guo},
	doi = {https://doi.org/10.1103/PhysRevA.82.042328},
	eprint = {1005.4510},
	journal = {Phys. Rev. A},
	pages = {042328},
	title = {Experimental investigation of the non-Markovian dynamics of classical and quantum correlations},
	url = {https://arxiv.org/pdf/1005.4510.pdf},
	volume = {82},
	year = {2010}}

@article{Xu_2013,
	author = {Xu, Jin-Shi and Sun, Kai and Li, Chuan-Feng and Xu, Xiao-Ye and Guo, Guang-Can and Andersson, Erika and Lo Franco, Rosario and Compagno, Giuseppe},
	doi = {10.1038/ncomms3851},
	issn = {2041-1723},
	journal = {Nature Communications},
	month = nov,
	number = {1},
	publisher = {Springer Science and Business Media LLC},
	title = {Experimental recovery of quantum correlations in absence of system-environment back-action},
	url = {http://dx.doi.org/10.1038/ncomms3851},
	volume = {4},
	year = {2013}}

@article{Franco:2012aa,
	author = {R. Lo Franco},
	doi = {10.1103/PhysRevA.85.032318},
	journal = {Physical Review A},
	number = {3},
	title = {Revival of quantum correlations without system-environment back-action},
	url = {https://doi.org/10.1103/PhysRevA.85.032318},
	volume = {85},
	year = {2012}}

@article{Monteiro_2025,
	author = {Monteiro, Luiz H. A.},
	doi = {10.3390/complexities1010002},
	issn = {3042-6448},
	journal = {Complexities},
	month = jun,
	number = {1},
	pages = {2},
	publisher = {MDPI AG},
	title = {A Simple Overview of Complex Systems and Complexity Measures},
	url = {http://dx.doi.org/10.3390/complexities1010002},
	volume = {1},
	year = {2025}}

@article{Rivat2024,
	author = {Rivat, S{\'e}bastien},
	day = {23},
	doi = {10.1007/s11229-024-04817-3},
	issn = {1573-0964},
	journal = {Synthese},
	month = {Dec},
	number = {1},
	pages = {11},
	title = {Open systems across scales},
	url = {https://link.springer.com/content/pdf/10.1007/s11229-024-04817-3.pdf},
	volume = {205},
	year = {2024}}

@article{Estrada2024,
	author = {Estrada, Ernesto},
	day = {01},
	doi = {10.1007/s10699-023-09917-w},
	issn = {1572-8471},
	journal = {Foundations of Science},
	month = {Dec},
	number = {4},
	pages = {1143--1170},
	title = {What is a Complex System, After All?},
	url = {https://link.springer.com/content/pdf/10.1007/s10699-023-09917-w.pdf},
	volume = {29},
	year = {2024}}

@article{Aolita:2015aa,
	author = {Leandro Aolita and Fernando de Melo and Luiz Davidovich},
	doi = {https://doi.org/10.1088/0034-4885/78/4/042001},
	eprint = {1402.3713},
	journal = {Reports on Progress in Physics},
	pages = {042001},
	title = {Open-System Dynamics of Entanglement},
	url = {https://arxiv.org/pdf/1402.3713.pdf},
	volume = {78},
	year = {2015}}

@article{Liu:2023aa,
	author = {Chang Liu and Haifeng Tang and Hui Zhai},
	doi = {https://doi.org/10.1103/PhysRevResearch.5.033085},
	eprint = {2207.13603},
	journal = {Physical Review Research,},
	pages = {033085},
	title = {Krylov Complexity in Open Quantum Systems},
	url = {https://arxiv.org/pdf/2207.13603.pdf},
	volume = {5},
	year = {2023}}

@article{Bhattacharyya:2022aa,
	author = {Arpan Bhattacharyya},
	doi = {10.1103/PhysRevD.105.046011},
	journal = {Physical Review D},
	number = {4},
	title = {Complexity for an open quantum system},
	url = {https://doi.org/10.1103/PhysRevD.105.046011},
	volume = {105},
	year = {2022}}

@book{Wiseman2010,
	author = {H. M. Wiseman and G. J. Milburn},
	number = {9780521804424},
	publisher = {Cambridge University Press},
	title = {Quantum Measurement and Control},
	year = {2010}}

@article{Garrahan:2010aa,
	author = {Juan P. Garrahan},
	doi = {10.1103/PhysRevLett.104.160601},
	journal = {Physical Review Letters},
	number = {16},
	title = {Thermodynamics of Quantum Jump Trajectories},
	url = {https://doi.org/10.1103/PhysRevLett.104.160601},
	volume = {104},
	year = {2010}}

@article{Manzano:2019aa,
	author = {Gonzalo Manzano},
	doi = {10.1103/PhysRevLett.122.220602},
	journal = {Physical Review Letters},
	number = {22},
	title = {Quantum Martingale Theory and Entropy Production},
	url = {https://doi.org/10.1103/PhysRevLett.122.220602},
	volume = {122},
	year = {2019}}

@article{Wiseman_1996,
	author = {Wiseman, H M},
	doi = {10.1088/1355-5111/8/1/015},
	issn = {1361-6625},
	journal = {Quantum and Semiclassical Optics: Journal of the European Optical Society Part B},
	month = feb,
	number = {1},
	pages = {205--222},
	publisher = {IOP Publishing},
	title = {Quantum trajectories and quantum measurement theory},
	url = {http://dx.doi.org/10.1088/1355-5111/8/1/015},
	volume = {8},
	year = {1996}}

@article{Daley_2014,
	author = {Daley, Andrew J.},
	doi = {10.1080/00018732.2014.933502},
	issn = {1460-6976},
	journal = {Advances in Physics},
	month = mar,
	number = {2},
	pages = {77--149},
	publisher = {Informa UK Limited},
	title = {Quantum trajectories and open many-body quantum systems},
	url = {http://dx.doi.org/10.1080/00018732.2014.933502},
	volume = {63},
	year = {2014}}

@article{Gneiting:2021aa,
	author = {Clemens Gneiting},
	doi = {10.1103/PhysRevA.104.062212},
	journal = {Physical Review A},
	number = {6},
	title = {Jump-time unraveling of Markovian open quantum systems},
	url = {https://doi.org/10.1103/PhysRevA.104.062212},
	volume = {104},
	year = {2021}}

@article{Lesanovsky:2013aa,
	author = {Igor Lesanovsky},
	doi = {10.1103/PhysRevLett.110.150401},
	journal = {Physical Review Letters},
	number = {15},
	title = {Characterization of Dynamical Phase Transitions in Quantum Jump Trajectories Beyond the Properties of the Stationary State},
	url = {https://doi.org/10.1103/PhysRevLett.110.150401},
	volume = {110},
	year = {2013}}

@article{Radaelli:2024aa,
	author = {Marco Radaelli},
	doi = {10.1103/PhysRevA.110.062212},
	journal = {Physical Review A},
	number = {6},
	title = {Gillespie algorithm for quantum jump trajectories},
	url = {https://doi.org/10.1103/PhysRevA.110.062212},
	volume = {110},
	year = {2024}}

@article{Xu:2019aa,
	author = {Zhenyu Xu},
	doi = {10.1103/PhysRevLett.122.014103},
	journal = {Physical Review Letters},
	number = {1},
	title = {Extreme Decoherence and Quantum Chaos},
	url = {https://doi.org/10.1103/PhysRevLett.122.014103},
	volume = {122},
	year = {2019}}

@article{Buca2019,
	author = {Bu{\v c}a, Berislav and Tindall, Joseph and Jaksch, Dieter},
	doi = {10.1038/s41467-019-09757-y},
	issn = {2041-1723},
	journal = {Nature Communications},
	month = apr,
	number = {1},
	publisher = {Springer Science and Business Media LLC},
	title = {Non-stationary coherent quantum many-body dynamics through dissipation},
	url = {http://dx.doi.org/10.1038/s41467-019-09757-y},
	volume = {10},
	year = {2019}}

@article{Dalibard1992,
	author = {Jean Dalibard},
	doi = {10.1103/PhysRevLett.68.580},
	journal = {Physical Review Letters},
	number = {5},
	pages = {580--583},
	title = {Wave-function approach to dissipative processes in quantum optics},
	url = {https://doi.org/10.1103/PhysRevLett.68.580},
	volume = {68},
	year = {1992}}

@article{Lempel_1976,
	author = {Lempel, A. and Ziv, J.},
	doi = {10.1109/tit.1976.1055501},
	issn = {1557-9654},
	journal = {IEEE Transactions on Information Theory},
	month = jan,
	number = {1},
	pages = {75--81},
	publisher = {Institute of Electrical and Electronics Engineers (IEEE)},
	title = {On the Complexity of Finite Sequences},
	url = {http://dx.doi.org/10.1109/TIT.1976.1055501},
	volume = {22},
	year = {1976}}

@article{Ziv_1977,
	author = {Ziv, J. and Lempel, A.},
	doi = {10.1109/tit.1977.1055714},
	issn = {1557-9654},
	journal = {IEEE Transactions on Information Theory},
	month = may,
	number = {3},
	pages = {337--343},
	publisher = {Institute of Electrical and Electronics Engineers (IEEE)},
	title = {A universal algorithm for sequential data compression},
	url = {http://dx.doi.org/10.1109/TIT.1977.1055714},
	volume = {23},
	year = {1977}}

@article{Vu:2021aa,
	author = {Tan Van Vu},
	doi = {10.1103/PhysRevLett.127.190601},
	journal = {Physical Review Letters},
	number = {19},
	title = {Lower Bound on Irreversibility in Thermal Relaxation of Open Quantum Systems},
	url = {https://doi.org/10.1103/PhysRevLett.127.190601},
	volume = {127},
	year = {2021}}

@article{Kiss:2005aa,
	author = {Istv{\'a}n Z. Kiss},
	doi = {10.1103/PhysRevLett.94.248301},
	journal = {Physical Review Letters},
	number = {24},
	title = {Predicting Mutual Entrainment of Oscillators with Experiment-Based Phase Models},
	url = {https://doi.org/10.1103/PhysRevLett.94.248301},
	volume = {94},
	year = {2005}}

@article{McNamara:1989aa,
	author = {Bruce McNamara},
	doi = {10.1103/PhysRevA.39.4854},
	journal = {Physical Review A},
	number = {9},
	pages = {4854--4869},
	title = {Theory of stochastic resonance},
	url = {https://doi.org/10.1103/PhysRevA.39.4854},
	volume = {39},
	year = {1989}}

@book{Carmichael1993,
	author = {Carmichael, Howard},
	doi = {10.1007/978-3-540-47620-7},
	isbn = {9783540476207},
	issn = {0940-7677},
	journal = {Lecture Notes in Physics Monographs},
	publisher = {Springer Berlin Heidelberg},
	title = {An Open Systems Approach to Quantum Optics: Lectures Presented at the Universit{\'e} Libre de Bruxelles October 28 to November 4, 1991},
	url = {http://dx.doi.org/10.1007/978-3-540-47620-7},
	year = {1993}}

@article{Vicentini:2018aa,
	author = {Filippo Vicentini},
	doi = {10.1103/PhysRevA.97.013853},
	journal = {Physical Review A},
	number = {1},
	title = {Critical slowing down in driven-dissipative Bose-Hubbard lattices},
	url = {https://doi.org/10.1103/PhysRevA.97.013853},
	volume = {97},
	year = {2018}}

@article{Rose_2016,
	author = {Rose, Dominic C. and Macieszczak, Katarzyna and Lesanovsky, Igor and Garrahan, Juan P.},
	doi = {10.1103/physreve.94.052132},
	issn = {2470-0053},
	journal = {Physical Review E},
	month = nov,
	number = {5},
	publisher = {American Physical Society (APS)},
	title = {Metastability in an open quantum Ising model},
	url = {http://dx.doi.org/10.1103/PhysRevE.94.052132},
	volume = {94},
	year = {2016}}

@article{Young:2020aa,
	author = {Jeremy T. Young},
	doi = {10.1103/PhysRevX.10.011039},
	journal = {Physical Review X},
	number = {1},
	title = {Nonequilibrium Fixed Points of Coupled Ising Models},
	url = {https://doi.org/10.1103/PhysRevX.10.011039},
	volume = {10},
	year = {2020}}

@article{Garbe_2020,
	author = {Garbe, Louis and Wade, Peregrine and Minganti, Fabrizio and Shammah, Nathan and Felicetti, Simone and Nori, Franco},
	doi = {10.1038/s41598-020-69704-6},
	issn = {2045-2322},
	journal = {Scientific Reports},
	month = aug,
	number = {1},
	publisher = {Springer Science and Business Media LLC},
	title = {Dissipation-induced bistability in the two-photon Dicke model},
	url = {http://dx.doi.org/10.1038/s41598-020-69704-6},
	volume = {10},
	year = {2020}}

@article{Krimer:2019aa,
	author = {Dmitry O. Krimer},
	doi = {10.1103/PhysRevA.100.013855},
	journal = {Physical Review A},
	number = {1},
	title = {Critical phenomena and nonlinear dynamics in a spin ensemble strongly coupled to a cavity. I. Semiclassical approach},
	url = {https://doi.org/10.1103/PhysRevA.100.013855},
	volume = {100},
	year = {2019}}

\end{document}